\documentstyle[emulateapj,apjfonts,epsfig,rotating,graphicx]{article}
\slugcomment{}

\begin{document}

\title{Neutrinos from Accreting Neutron Stars}

\author{Luis A. Anchordoqui$^a$, Diego F. Torres$^b$, Thomas P.
McCauley$^a$, Gustavo E. Romero$^c$, \\
\& Felix A. Aharonian$^d$}

\affil{$^a$ Department of Physics, Northeastern University, 110
Forsyth St., Boston, MA 02115, USA.}
\affil{$^b$ Lawrence Livermore National Laboratory, 7000 East
Ave., L-413, Livermore, CA 94550, USA. E-mail:dtorres@igpp.ucllnl.org} 
\affil{$^c$ Instituto Argentino de
Radioastronom\'{\i}a (IAR), C.C.\ 5, 1894 Villa Elisa, Argentina}
\affil{$^d$ Max-Planck-Institut f\"ur Kernphysik, Posfach 10 39
80, D-69029 Heidelberg, Germany}

\begin{abstract}
The magnetospheres of accreting neutron stars develop
electrostatic gaps with huge potential drops. Protons and ions,
accelerated in these gaps along the dipolar magnetic field lines
to energies greater than 100 TeV can impact onto the surrounding
accretion disc. A proton-induced cascade develops, and charged
pion decays produce $\nu$-emission. With extensive disc shower
simulations using {\sc dpmjet} and {\sc geant4}, we have
calculated the resulting $\nu$-spectrum. We show that the spectrum
produced out of the proton beam is a power law. We use this result
to propose accretion-powered X-ray binaries (with highly
magnetized neutron stars) as a new population of point-like
$\nu$-sources for km-scale detectors such as ICECUBE. As a
particular example we discuss the case of A0535+26. We show that
ICECUBE should find A0535+26 to be a periodic $\nu$-source, one
for which the formation and loss of its accretion disc can be
fully detected. Finally, we briefly comment on the possibility
that smaller telescopes, like AMANDA, could also detect A0535+26
by folding observations with the orbital period.
\end{abstract}

\keywords{X-ray binaries: general, gamma$-$rays: theory,
gamma$-$rays: observations, neutrinos: observations, X-ray
binaries: A0535+26}

\section{Introduction}

X-ray binaries have fascinated those looking for Galactic neutrino
sources (e.g.~Berezinsky et al. 1985, Kolb et al. 1985, Gaisser
and Stanev 1985). The basic idea is to use somehow the secondary
object in the system to accelerate protons, which then could
collide within a higher density medium. One possibility that was
earlier explored was to use the primary itself as the target for
those accelerated hadrons~(Berezinsky et al. 1990). Depending on
the grazing angle of the colliding protons, on the size and type
of the primary star, and on the effectiveness of the acceleration
mechanism, $\gamma$-rays and neutrinos (or $\nu$s) could escape
from the system. $\gamma$-rays, however, could more naturally be
produced in the accretion disc surrounding the neutron stars;
detailed models for this possibility were presented by Cheng and
Ruderman (1989). Herein, we show that accreting X-ray binaries in
which the compact object is a magnetized neutron star are sources
of high energy neutrinos that can be detected by forthcoming
neutrino telescopes. Moreover, we show that the signal-to-noise
ratio can be high enough as to allow timing studies and
multiwavelength comparison.

\section{The accretion disc of A0535+26}

Of all the X-ray binaries, A0535+26 is one of the most studied.
A0535+26 is a Be/X-ray transient where the compact object is a 104
second pulsar in an eccentric orbit around the B0III star HDE
245770. Be stars are rapidly rotating objects that eject mass,
irregularly forming gaseous discs in their equatorial planes.
Strong and recurrent X-ray outbursts were observed separated by
111 days, which has been identified with the orbital period
(Giovannelli and Sabau Graziati 1992). These outbursts occur when
the accretion onto the neutron star increases at periastron
passage. (The average ratio of the X-ray luminosity at the
periastron to that of apoastron is $\sim~100$~, Janot-Pacheco et
al. 1987.)
 The BATSE instrument of the Compton Gamma Ray Observatory
detected a 33-day, broad, quasi-periodic oscillation in the power
spectra of the X-ray flux, definitively showing that an accretion
disc is formed during giant outbursts~(Finger et al. 1996). The
$\gamma$-ray light curve maximum for this object is
anti-correlated with X-rays at periastron passage (Hartman et al.
1999, Romero et al. 2001). To explain this anti-correlation, a
variation in the disc grammage is invoked: when the disc is fully
formed, and the X-ray luminosity is at its maximum, the disc
grammage is too high as to allow $\gamma$-ray photons above
100~MeV to escape. We now describe how the accretion disc model
works.

\begin{figure*}
\centering
\includegraphics[width=0.5\textwidth,height=7cm]{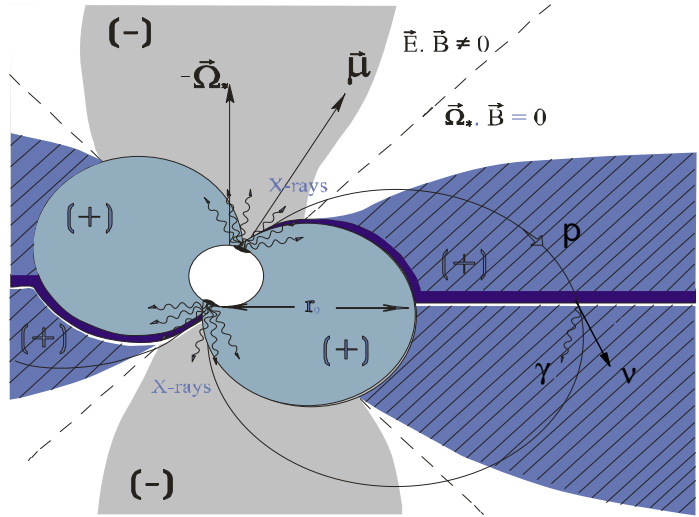}
\caption{Sketch (not to scale) of the magnetosphere of an
accreting magnetized neutron star, with the electrostatic gap
produced when $\Omega_{*}<\Omega_{\rm d }$. Protons entering into
the gap from the region that co-rotates with the star (-) are
accelerated along the field lines and impact onto the disc,
initiating a shower. $\nu$s and $\gamma$-ray photons, resulting
from pion decay, emerge from the opposite side of the disc and
will generically be beamed due to momentum conservation. This
figure is adapted from several of the works by Cheng and Ruderman,
quoted in the text. } \label{1}
\end{figure*}

%%%%%%%%%%%%%%%%%%% CHENG ET AL MODEL  %%%%%%%%%%%%%%%%%%%%%%%%%%%%%
Accretion discs can penetrate the stellar magnetospheres of
accreting rotating neutron stars ~(Ghosh and Lamb 1979). This
penetration creates a broad transition zone between the
unperturbed disc flow ---far from the star--- and the co-rotating
magnetospheric flow ---close to the star. In the transition zone,
with inner radius $r_0$, the angular velocity is Keplerian.
Between $r_0$ and the co-rotation radius $r_{\rm co}$ there is a
thin boundary layer where the angular velocity significantly
departs from the Keplerian value. At $r_{\rm co}$ the disc is
disrupted by the magnetic pressure and the accreting mass is
channelled by the field to impact onto the stellar surface,
producing strong X-ray emission.

 The magnetosphere of an
accreting neutron star spinning  slower than that of the accretion
disc can be divided into three regions: i) a region coupled to the
star by the magnetic field lines that do not penetrate the disc,
ii) an equatorial region linked to the disc by the field attached
to it, and iii) a gap entirely empty of plasma separating both
regions~(Cheng and Ruderman 1989). The magnetospheric region
penetrated by field lines which do not intersect the disc
co-rotates with the star at an angular velocity $\Omega_{*}$. When
the angular velocity of the disc $\Omega_{\rm d }$ exceeds the
angular velocity of the star ($\Omega_{*}<\Omega_{\rm d }$), the
equatorial plasma between the inner accretion disc radius $r_0$
and the Alfv\'en radius $r_{\rm A}$ co-rotates with the disc to
which it is linked by the frozen field lines. Inertial effects
result in a charge separation around the ``null surface", ${\bf
\Omega_{*}\cdot B}=0$. This leads to the formation of an
electrostatic gap, with no charge (see Fig.~1). In this gap ${\bf
E\cdot B}\neq 0$, and a strong potential drop is established. The
maximum potential drop along the magnetic field lines through the
gap is~(Cheng and Ruderman 1989)
\begin{equation}
\!\!V_{\rm max} \sim 4 \times 10^{14}
\beta^{-5/2}\left(\frac{M_{*}}{M_{\odot}}\right)^{1/7}\!\!
R_6^{-4/7}L_{37}^{5/7}B_{12}^{-3/7}\;\;{\rm V},
\end{equation}
where $M_{*} \sim 1.4-2.7$ $M_{\odot}$ is the neutron star mass
(Giovannelli and Sabau Graziati 1992), $R_6 \equiv R_*/10^{6}~{\rm
cm}$ is its radius, $B_{12} \equiv B_*/10^{12}~{\rm G}$ is the
magnetic field of the star, $L_{37}$ is the X-ray luminosity in
units of $10^{37}$ erg s$^{-1}$, and $\beta\equiv 2r_0/r_{\rm A}$
is twice the ratio between the inner accretion disc radius and the
Alfv\'en radius. The Alfv\'en radius for spherical accretion can
be determined from the condition that the unscreened magnetic
energy density of the stellar field becomes comparable to the
kinetic energy density of the accreting matter~(Cheng and Ruderman
1989),
\begin{equation}
r_{\rm A} \approx 3 \times 10^8 \, L_{37}^{-2/7}\, B_{12}^{4/7} \,
\left(\frac{M_*}{M_\odot}\right)^{1/7}\, R_6^{10/7} {\rm cm}.
\end{equation}
The electrostatic gap hovers on the accretion disc from the
innermost disc radius, $\sim r_0$ up to a distance of about
$r_{\rm A}$. Without loss of generality we can assume that the
stellar magnetic dipolar moment ($\mu = B_* R_*^3/2$) that induces
$B$--field lines across the disc is aligned with the rotation axis
of the system~(Cheng and Ruderman 1991). The $B$--field on the
disc, however, strongly depends on the screening factor generated
by currents induced in the disc surface. In what follows, we take
a mean $B$--field of 6000~G~(Ghosh and Lamb 1979b).

Protons entering into the gap from the stellar co-rotating region
are accelerated up to energies $ e\,V_{\rm max}$ and directed to
the accretion disc by the field lines. The maximum current that
can flow through the gap can be determined from the requirement
that the azimuthal magnetic field induced by the current does not
exceed that of the initial magnetic field~(Cheng and Ruderman
1989)
\begin{equation}
J_{\rm max} \sim 1.5\times10^{24}
\beta^{-2}\left(\frac{M_*}{M_{\odot}}\right)^{- 2/7}
\!\!\!\!\!R_6^{1/7}L_{37}^{4/7}B_{12}^{-1/7}\;{\rm esu\;s^{-1}}
\,.
\end{equation}
\mbox{} The mean number of protons impacting the disc is huge, $
N_p  = J_{\rm max}/e \sim 10^{33}$~s$^{-1}$, and the total power
deposited by the proton beam in the disc is $ P_{\rm max}= J_{\rm
max} V_{\rm max} \sim 10^{36}$~erg~s$^{-1}$. The collision of the
relativistic proton beam with the disc initiates hadronic and
electromagnetic showers, in which high energy $\nu$s are produced
from  decay of charged pions, whereas $\gamma$-rays are produced
from the decay of their neutral partners.

%%%%%%%%%%%%%%%%%%% CHENG ET AL MODEL  %%%%%%%%%%%%%%%%%%%%%%%%%%%%%

%%%%%%%%%%%%%%%%%%% SYSTEM  %%%%%%%%%%%%%%%%%%%%%%%%%%%%%
\section{Phenomenology}

{\it The system just described is an ideal source of high energy
$\nu$s: It has a very dense material for $pp$--interactions, and,
at the same time, the region of acceleration is separated from
that of high density, where interactions occur.} The high energy
$\nu$-production in accretion discs of neutron stars is, however,
subject to very stringent conditions on the disc grammage. The
latter should be large enough as to allow protons to interact, and
small enough as to allow pions to decay or to escape, so as to
avoid losing energy. But the disc grammage is actually a
periodically varying function in these systems, which follows the
orbital dynamics. As noted, this fact was used to explain why
X-ray maximum is coincident with a non-detection in the
$\gamma$-ray band for the system A0535+26 (Romero et al. 2001). An
examination of the cross sections involved is in order.

%%%%%%%%%%%%%%%%%%% CROSS SECTIONS %%%%%%%%%%%%%%%%%%%%%%%%%%%%%

In the energy range of interest, the $p \pi$ cross section,
$\sigma_{p\pi} \sim 25$~mb~(Carroll et al. 1979), is not too far
away from the cross section for $\gamma$-ray absorption in the
Coulomb field produced by disc ions, $\sigma_{\gamma E } \,\sim
10~{\rm mb}$~(Cox et al. 2001, Bethe and Heitler 1934). Then,
photons with energies $> 100$~MeV undergo interactions within
typical accretion discs if the mean density of hydrogen is $n
\gtrsim 1 \times 10^{20}$~cm$^{-3}$ (the survival probability for
$\gamma$'s to a distance comparable to the thickness of the disc
due to this process is less than 0.1\%). Moreover, assuming a
source luminosity $L_{37} \sim 1$ in the form of $\approx 1$~keV
photons and an average disc radius of $\overline{r}  \approx 2
\times  10^8$~cm, the energy density inside the disc is roughly
$L_{37}/(2 \pi\, \overline{r}^2 \,c) \simeq 10^9\,\, {\rm
erg/cm}^3$, implying a number density of 1 keV photons $\sim
10^{18}~{\rm cm}^{-3}$.
%On the other hand, the energy density of 6000~G magnetic field ($B^2/(8 \pi)$)
%is 3 orders of magnitude less. Therefore, we can ignore the synchrotron losses,
%because
Therefore, the secondary electrons and $\gamma$-rays most
effectively interact with the radiation field of the disc. These
secondary electrons originating from decays of neutral and charged
pi-mesons will trigger an electromagnetic cascade in this field
resulting to a standard cascade spectrum with a cutoff energy
around $m_e^2 c^4/\epsilon \approx 250 \,\, (\epsilon/1 {\rm
keV})^{-1} {\rm MeV}$, where $m_e$ is the mass of the electron and
$\epsilon$ is the average energy of the background thermal
photons.

%%%%%%%%%%%%%%%%%%% DISC  %%%%%%%%%%%%%%%%%%%%%%%%%%%%%

According to canonical accretion disc models, the average density
scales as  $n  \propto \,L_{37}^{20/35} \,
(M_*/M_\odot)^{-1/28}\,\, r^{-21/20}$~(Shapiro and Teukolsky
1983). We normalize the particle density of the accretion disc at
$\overline{r}$  in order to fit $\gamma$-ray observations of
A0535+26 (strong photon absorption in the periastron of the
system, where $L_{37} \approx 1$~, Finger et al. 1996). Because of
the similarity in the cross sections discussed above, whenever
$\gamma$-rays are absorbed, there will be a much reduced signal in
$\nu$s. Otherwise stated, the hadronic shower will also cool down
to sub-100 GeV energies. But as the disc density is a function of
the position in the orbit, before and after periastron passage,
there are orbital configurations prone to the emission of both,
high energy photons and neutrinos.

To simplify the discussion and get a numerical example of the
processes involved, we set the half thickness of the disc to a
constant ($h \sim 3.5 \times 10^6$~cm) and consider an orbital
configuration for which $L_{37} \sim 0.2$, before or after
periastron passage. The cross section for $pp$ interactions is
$\sim 50$ mb ~(Battiston et al. 1992). Then, the probability for a
$pp$ interaction to take place within the disc is $1 - e^{-2h
/\lambda_p} \gg 99\%$, where $\lambda_p$ is the mean free path of
the relativistic protons. It is worthwhile to point out that even
though the photon and particle densities are similar, the $p
\gamma$ cross section is two order of magnitude smaller than the
$pp$ cross section.

On the other hand, the pion mean free path in this configuration
is $\lambda_\pi \sim 10^6$~cm, which implies a charged pion
survival probability ---to its decay distance--- of $ P \approx
e^{-\Gamma\,c\,\tau_\pi/\lambda_\pi}\,\,,$ where $\tau_\pi$ is the
$\pi^\pm$ lifetime, and $\Gamma$ is its Lorentz factor. Then, the
probability for producing $\nu$s of energy: 200 GeV and 400 GeV is
---on average---  10\% and 1\%, respectively. We expect a rapidly
fading signal of neutrinos with energies greater than 1 TeV.

There appears to be enough room for the disc density (or grammage)
to acquire values such that all protons will partake in hadronic
interactions, whereas some of the pions produced will decay to
high energy neutrinos at a level significant enough as to show up
in neutrino telescopes. To analyze this in detail, numerical
simulations of a proton-induced cascade unfolding within an
accretion disc were performed.

%%%%%%%%%%%%%%%%%%%%%%%%%%%%% SIMULATION: HARD %%%%%%%%%%%%%%%%%%%%%%%%

\section{Numerical Simulations}

\begin{figure*}
\centering
\includegraphics[width=0.6\textwidth,height=9.5cm]{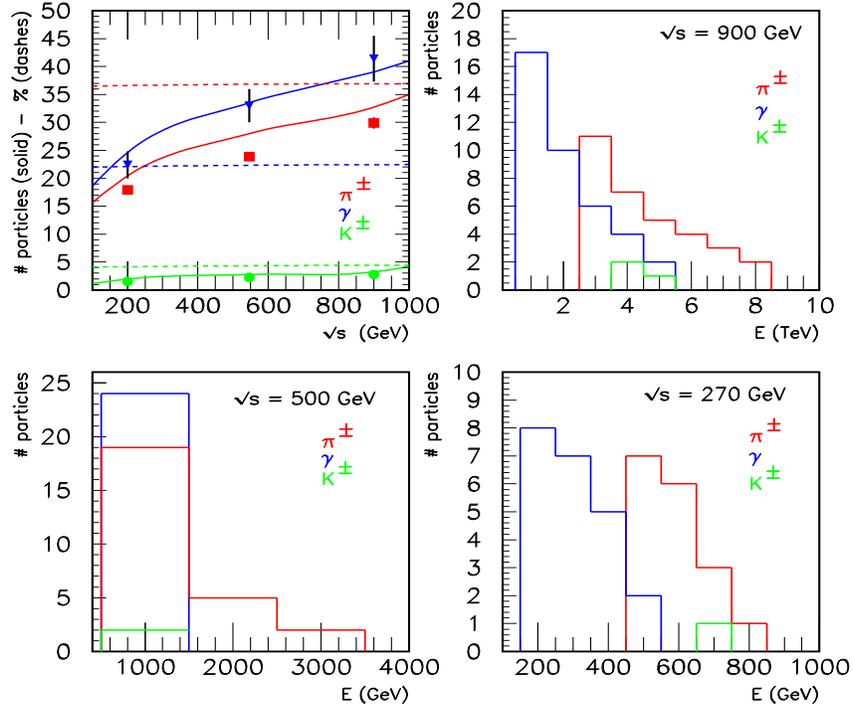}
\caption{Characteristics of hadronic interactions implemented in
the Monte Carlo event generator {\sc dpmjet}--II. In the first
panel we show the average multiplicities from $p N$ collisions as
a function of the center-of-mass energy (solid lines) together
with the energy fraction (in $\%$) going into secondary particles
(dashed lines), from bottom to top, $K^\pm$, $\pi^\pm$, and
$\gamma$. The circles, squares, and triangles give the average
numbers of secondary particles produced in non-diffractive $p \bar
p$ collisions measured by the $UA5$ Collaboration~(Ansorge et al.
1989). The other three panels show the spectra of secondary
particles produce in $p N$ collisions at different center-of-mass
energies. The spectrum of $K^{0L} + K^{0S}$ is roughly equal to
that of $K^{\pm}$.} \label{5}
\end{figure*}

\begin{figure*}
\centering
\includegraphics[width=0.5\textwidth,height=8.5cm]{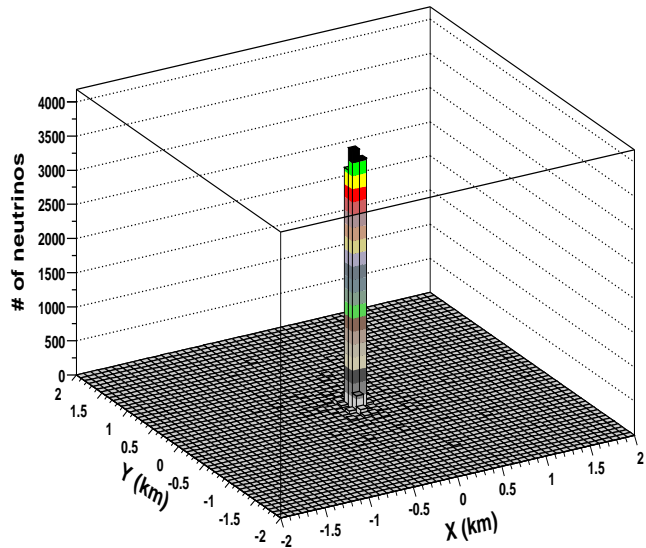}\hspace{0.6cm}
\caption{Neutrino lateral distribution from a proton shower in a
typical accretion disc setting. The $xy$-plane is parallel to the
disc. Because the secondary pions are produced with very low
transverse momentum, most neutrinos are produced near the shower
core.} \label{mi}
\end{figure*}

The accretion disc itself is simulated as a cylindrical volume of
thickness $2h$ and is filled with elemental hydrogen at a density
$n = 4 \times 10^{19}$~cm$^{-3}$.  A 400~TeV value is retained as
the average energy of each proton in the beam even for $L_{37}
\approx 0.2$, because the uncertainties in all other parameters
involved dilute the deviation from the fiducial result. The
generation and tracking of secondary particles in the cascade
development was performed using {\sc geant4}: a simulation toolkit
(designed for operation up to center-of-mass energies $\sqrt{s}
\sim 100$ GeV) that provides general-purpose tools for the
simulation of the passage of particles through matter~(Agostinelli
et al. 2002). We process the initial hadronic collisions (with
$\sqrt{s}
> 100$~GeV) using the event generator {\sc dpmjet}-II~(Ranft
1995). This program, based on the Gribov--Regge theory, describes
soft particle interactions by the exchange of one or multiple
Pomerons (the inelastic reactions are simulated by cutting
Pomerons into color strings which subsequently fragment into color
neutral hadrons). In the energy range of interest, $\sqrt{s} \sim
10^2 - 10^{3}~{\rm GeV}$, the average energy fraction of the
highest energy baryon in the simulated $pN$ collisions is roughly
30\%. Therefore, on average, after three collisions, the energy of
the leading particle will be degraded down to $\sim$10~TeV. The
secondary meson spectra from these three collisions are given in
Fig.~\ref{5}. Additionally, for these center-of-mass energies, the
soft baryon channel comprises 2 nucleons, each carrying (on
average) 1\% of the energy of the incoming proton. These secondary
distributions together with the leading particle ---degraded in
energy--- were injected into the cylindrical volume representing
the disc. The position of the first interaction was selected
randomly following a Poisson distribution with a mean equal to the
proton mean free path. The second and third interaction points
were selected with the same procedure, taking the preceding
collision as reference, and going further down within the disc. We
then use the {\sc geant4} to track all particles. In this second
step, all hadronic collisions are processed with {\sc geant4}
implementation of {\sc geisha}~(Fesefeldt et al. 1985), a program
tuned to analyze experimental results (of a variety of projectiles
and targets) in the few-GeV energy range. A magnetic field of
6000~G threads the synthetic disc volume, whose direction is
pointing upwards in Fig.~1. Variations in the $B$-field across the
disc do not affect the results presented below, because most of
the particles in the shower are produced in ``minimum bias''
events boosted in the forward direction (see Fig.~\ref{mi}).

%%%%%%%%%%%%%%%%%%%%%%%%%%%%% SIMULATION %%%%%%%%%%%%%%%%%%%%%%%%

One can see in  Fig.~\ref{2} the average number of $\nu$s produced
from pion decay per unit energy (bin size $\equiv \Delta E_\nu =
50$ GeV). The tail of the energy distribution is very well fitted
by a power-law, $A_0\,E_\nu^{-\gamma}$, with values of $A_0$ and
$\gamma$ given in the figure. {\it Therefore, a monoenergetic beam
of high energy protons impacting onto an accretion disc produces
outgoing $\nu$s with a power law spectrum.}

\section{The neutrino signal}

Having estimated the $\nu$-spectrum, we now analyze the
signal-to-noise (S/N) ratio for a km-scale detector like
ICECUBE~(Karle et al. 2002). This detector will consist of 4800
photomultipliers, arranged on 80 strings placed at depths between
1400 and 2400 m under the South Pole ice. The strings will be
located in a regular space grid covering a surface area of 1
km$^2$. Each string will have 60 optical modules (OM) spaced 17 m
apart. The number of OMs which have seen at least one photon
(generated by \v{C}erenkov radiation produced by the muon which
resulted from the interaction of the incoming $\nu$ in the Earth's
crust) is called the channel multiplicity, $N_{\rm ch}$. The
multiplicity threshold is set to $N_{\rm ch}=5$,which corresponds
to an energy threshold of 200~GeV~(Alvarez-Mu\~niz and Halzen
2002). The angular resolution of ICECUBE will be 0.7$^\circ$,
which implies a search window of $\sim$1$^\circ$ radius~(Karle et
al. 2002).

\begin{figure*}
\centering
\includegraphics[width=0.4\textwidth,height=8.5cm]{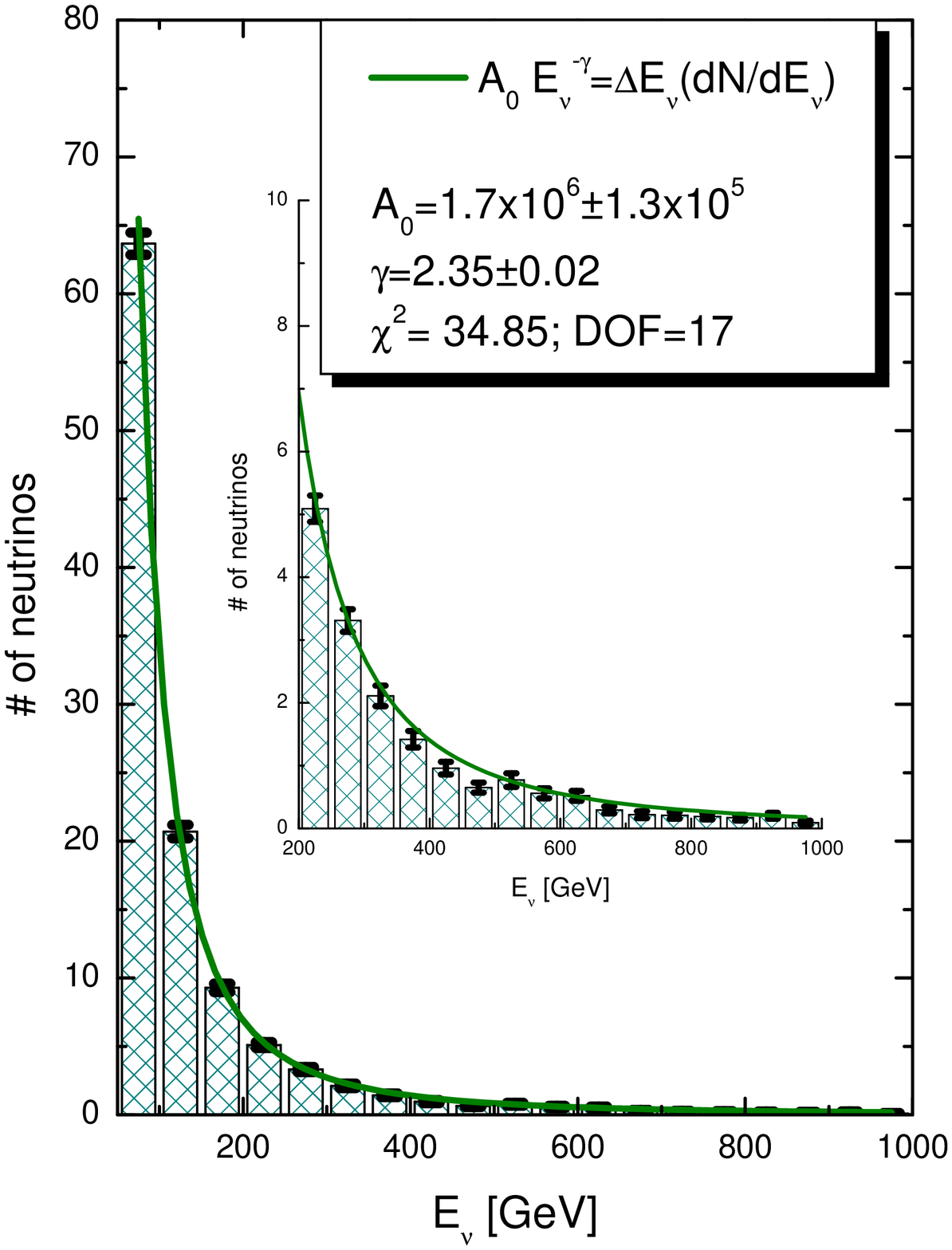}\\\vspace{.5cm}
\caption{Neutrino energy distribution generated by pion decay
induced by the collision of 400 TeV protons onto an accretion disc
of a typical X-ray binary. The error bars indicate the RMS
fluctuations for each of the mean values, represented by the
height of each box in the histogram. The latter were obtained
averaging over 100 showers. The solid line is a fit to this
spectrum, whose parameters are shown in the insert. The inset
figure is the high energy tail of the distribution, together with
the same fit.} \label{2}
\end{figure*}

The event rate of the atmospheric $\nu$-background that will be
detected in the search bin is given by
\begin{equation}
\left. \frac{dN}{dt}\right|_{\rm B} = A_{\rm eff}\, \int dE_\nu
\,\frac{d\Phi_{\rm B}}{dE_\nu}\, P_{\nu \to \mu}(E_\nu)\,\,\Delta
\Omega_{1^\circ \times 1^\circ}\,, \label{background}
\end{equation}
where $A_{\rm eff}$ is the effective area of the detector, $\Delta
\Omega_{1^\circ \times 1^\circ} \approx 3 \times 10^{-4}$~sr,
 and
$d\Phi_{\rm B}/dE_\nu \lesssim 0.2 \,(E_\nu/{\rm
GeV})^{-3.21}$~GeV$^{-1}$ cm$^{-2}$ s$^{-1}$ sr$^{-1}$  is the
$\nu_\mu + \bar \nu_\mu$ atmospheric $\nu$-flux~(Volkova 1980,
Lipari 1993). Here, $P_{\nu \to \mu} (E_\nu) \approx 3.3 \times
10^{-13} \,(E_\nu/{\rm GeV)}^{2.2}$ denotes the probability that a
$\nu$ of energy $E_\nu \sim 1 - 10^3~{\rm GeV}$, on a trajectory
through the detector, produces a muon~(Gaisser et al. 1995). On
the other hand, the $\nu$-signal is
\begin{equation}
\left. \frac{dN}{dt}\right|_{\rm S} =  A_{\rm eff} \,\int dE_\nu
\, \frac{d\Phi_{\nu_\mu}}{dE_\nu}\,P_{\nu \to \mu}(E_\nu)\  \,,
\label{yellowsubmarine}
\end{equation}
where
\begin{equation}
\frac{d\Phi}{dE_\nu} = \left(\frac{\Delta
\Omega}{4\,\pi}\right)^{-1} \frac{1}{4\, \pi\, d^2}\,
\frac{d\Phi_0}{dE_\nu}\, \label{osc}
\end{equation}
is the incoming $\nu$-flux, emitted by a source at a distance $d$
from Earth with beaming factor $\Delta \Omega/4\pi$.
Phenomenological estimates yield $\Delta \Omega / 4\pi \sim
0.1$~(Cheng and Ruderman 1989). The $\nu$-emission spectrum is
just ${d\Phi_0}/{dE_\nu} = N_p \, dN/dE_{\nu}$, where
$dN/dE_{\nu}$ can be read from the first panel of Fig.~\ref{2},
and $N_p \approx 3 \times 10^{33}$ is the total number of protons
impacting onto the disc. It is noteworthy that the flux of
neutrinos, which is dominantly $\nu_\mu + \bar{\nu}_\mu$ at
production, is expected to be completely mixed in flavor upon
arrival at Earth. Specifically, there is now strong evidence for
maximal mixing among all neutrino species~(Fukuda et al. 1998),
which implies that the $\nu$-flux would be completely mixed after
a propagation distance of~(Bilenky et al. 1999)
\begin{equation}
L_{\rm osc} \approx 2.5 \, \frac{E_\nu}{{\rm GeV}} \,\frac{{\rm
eV}^2}{\Delta m_\nu ^2}\,\,\,{\rm km}\,.
\end{equation}
where $\Delta m_\nu^2 \gtrsim 10^{-6}$~eV$^2$ is the $\nu$--mass
splitting. Putting all this together we obtain
\begin{equation}
\frac{d\Phi_{\nu_\mu}}{dE_\nu} \approx {1\over 3}
\frac{d\Phi}{dE_\nu}\,\,.
\end{equation}

\begin{figure*}
\centering
\includegraphics[width=0.6\textwidth,height=10cm]{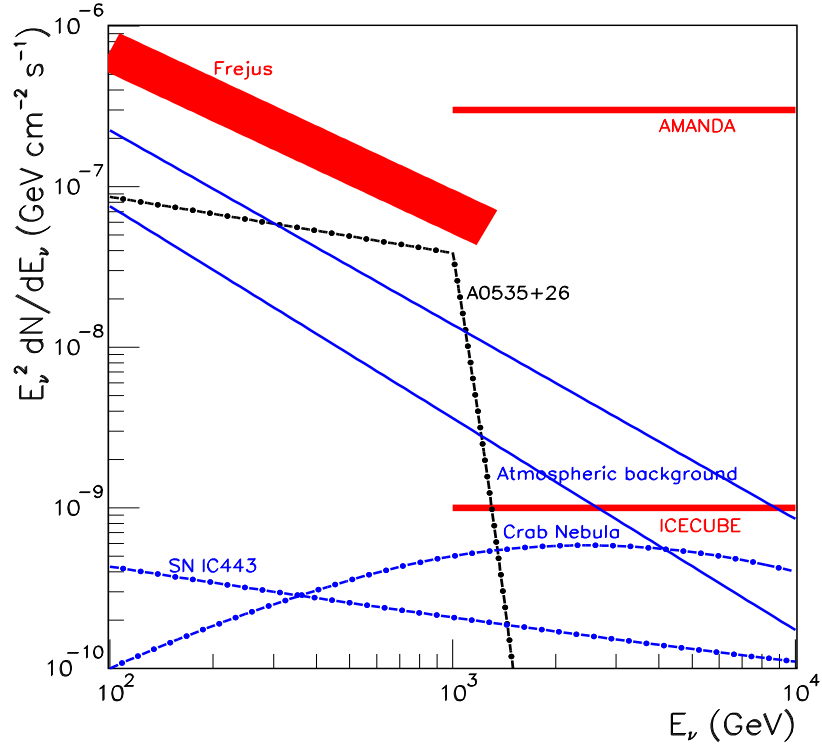}
\caption{The expected $\nu_\mu + \bar{\nu}_\mu$--flux coming from
A0535+26 is represented by the decreasing dash-dotted curve with
sharp cutoff above 1~TeV. The flux above this energy is subject to
large uncertainties because of shower-to-shower fluctuations, and
it is not being taken into account to compute the S/N ratio, which
is based only in the interval (300 GeV -- 1~TeV). The shape of the
cutoff given in this figure corresponds to that estimated using
the phenomenological arguments about pion survival probability
discussed in the main text. The atmospheric $\nu_\mu +
\bar{\nu}_\mu$-background within a $1^\circ \times 1^\circ$ square
of the source is indicated by the two diagonal solid lines, which
corresponds to horizontal and vertical (upper curve)
impacts~(Volkova 1980, Lipari 1993). Also shown is the current
upper limit on $\nu_\mu + \bar{\nu}_\mu$--flux for a source with
differential energy spectrum $\propto E^{-2}$ and declination
angle larger than 40$^\circ$ as reported by the AMANDA
Collaboration (Ahrens et al. 2002), as well as the expected
sensitivity of ICECUBE after its first three years of
operation~(Hill et al. 2001). Assuming the flux is fairly uniform
with decreasing angular bin, we scaled down to $\Delta
\Omega_{2^\circ \times 2^\circ}$ (angular resolution of AMANDA)
the current upper limit on diffuse $\nu_\mu +
\bar{\nu}_\mu$--fluxes (90\% CL)  as reported by the Fr\'ejus
Collaboration~(Rhode et al. 1996). We stress that the Fr\'ejus
detector was located in  an underground laboratory in the Alps and
consequently it had not a complete field of view of A0535+26. The
decreasing dash-dotted line indicates the $\nu_\mu +
\bar{\nu}_\mu$-spectrum of SN IC443~(Hill et al. 2001). This curve
relies on the assumption that (on average) a pair of muon
neutrinos with energy $E_\pi/4$ are produced in the
pion-to-muon-to-electron decay chain of proton showers (in very
low density media) for every photon with energy $E_\pi/2$~(Gaisser
et al. 1998). The bumpy curve indicates one of the hypotesyses
(Model-II in Bednarek and Protheroe 1997) for $\nu_\mu +
\bar{\nu}_\mu$-production in the Crab Nebula.} \label{3}
\end{figure*}

A0535+26 is very close to Earth, distant only $\sim
2.6$~kpc~(Giovannelli and Sabau Graziati 1992). In addition, as
only $\nu$s going through the Earth can be identified, the
northern location of this source makes it accessible from the
south pole site of ICECUBE. It can be seen in Figure~\ref{3} the
expected $\nu$-flux from A0535+26. The atmospheric
$\nu$-background for a $1^\circ \times 1^\circ$ bin is indicated
by the diagonal lines. Also indicated in the figure are the fluxes
of several others point sources. It is clear that for $300~{\rm
GeV} < E_\nu < 1~{\rm TeV}$, $\nu$s coming from A0535+26 can
dominate both over the background and over all other known
sources. The S/N ratio in the range $E_\nu \in [300\, {\rm GeV},
1\, {\rm TeV}]$, using Eqs.~(\ref{background}) and
(\ref{yellowsubmarine}) and considering an observing time of $\sim
50$~days, compatible with the period in which the accretion disc
is forming or disappearing, is
\begin{equation}
\frac SN = \frac{N|_{_{\rm S}}}{\sqrt{N|_{_{\rm B}}}} \approx 1.7
\, . \label{SN}
\end{equation}
It is possible, then, that A0535+26 could be detected as a
variable neutrino source even within the time span of one orbital
period. The possibility of detection can be further improved by
adding up several orbital periods, of which there are three per
year. It is therefore possible that sources like A0535+26 could
have been already detected in the AMANDA data. In order to search
for this possibility two all-sky maps could be used. \footnote{We
thank Rodin Porrata (LLNL) for bringing this to our attention}
One, adding up all observations in periods where the source is
expected to be ON, before and after periastron passage. Another,
adding up all observations with the source in OFF mode, at the
periastron. The difference between these two maps, weighted by the
respective integration times, should leave only one source on the
sky, disregarding that this source in itself is well below the
background for the experiment. Of course, the same technique can
work for ICECUBE. A few years integration time would add up tens
of orbital revolutions, and detection of A0535+26 should be
unambiguous. Indeed, A0535+26 is, in projection, less than
5$^\circ$ away from the Crab Nebula, which was suggested as a
potential detection by AMANDA (Barwick et al. 2002). This
detection is based on a $\sim 6^\circ\times 6^\circ$ binning of
the nothern hemisphere. It is not implausible, then, that A0535+26
may significantly contribute to this angular bin, particularly at
energies below 1~TeV.

From a technical point of view, ICECUBE must select a soft energy
cut-off to observe this source, since A0535+26 shines mostly below
1 TeV (the channel mutiplicity threshold of the experiment must be
set as $5\leq N_{\rm ch} \leq 30$~(Karle et al. 2002). A next
generation observatory, like NEMO~(Riccobene et al. 2002), would
improve the angular resolution to $\sim 0.3$$^\circ$, which would
increase the $S/N$-ratio  by a factor of three.

\hfill

\section{Concluding remarks}

In summary, a wide range of parameters able to describe the
physical situation of accreting neutron stars make them a
bona-fide population of $\nu$-emitters with power-law spectra. In
the particular of case of A0535+26, the $\nu_\mu +
\bar{\nu}_\mu$--flux between 300~GeV and 1~TeV overwhelm those of
all other point sources, and is clearly above the atmospheric
background in a $1^\circ \times 1^\circ$ angular bin.
%It is so much above background that the S/N ratio can distinguish
%the position in the binary orbit:
% $\nu$-observations of this X-ray binary can map the
%structure, formation, and loss of its accretion disc.
The neutrino signal will be periodic in nature. The accumulated
signal during the apoastron --where there is no disc formed-- and
periastron passages will always be at the noise level. On the
contrary, the accumulated signal during time spans before and
after the periastron passage, will be, as we have shown,
significantly different from noise level. A reasonable integration
time for neutrinos would then secure high-confidence detections of
the appearance and disappearance of the accretion disc ($\gtrsim
3$yr to reach a 5$\sigma$ effect). Complementary information from
high-energy photon astronomy in the X- and $\gamma$-ray domain
will be essential to confirm the hadronic origin of the radiation
in accreting neutron star systems, as well as to study the
formation and loss of accretion discs in eccentric binary systems.

\acknowledgments We have benefitted from discussions with Haim
Goldberg, Christopher Mauche, and Rodin Porrata. The research of
L.A.A. and T.P.M. was supported by the U.S. National Science
Foundation, Grant N$^o$ PHY-0140407. The work of D.F.T. was
performed under the auspices of the U.S. Department of Energy by
University of California Lawrence Livermore National Laboratory
under contract No. W-7405-Eng-48. D.F.T. is Lawrence Fellow in
Astrophysics. The research of G.E.R. is mainly supported by
Fundaci\'on Antorchas, with additional contributions from the
agencies CONICET (PIP N$^o$ 0430/98) and ANPCT (PICT 03-04881). He
is a member of CONICET.


\begin{thebibliography}{99}

\bibitem{Alvarez-Muniz:2002tn}
Alvarez-Mu\~niz J. and Halzen F. 2002,
%``High-energy neutrinos from the cosmic accelerator RX J1713.7-3946,''
Astrophys.\ J.\  576, L33.
%%CITATION = ASTRO-PH 0205408;%%


\bibitem{Agostinelli:2002hh}
Agostinelli S. et al.  [{\sc geant4} Collaboration] 2002,
%``GEANT4: A simulation toolkit,''
SLAC-PUB-9350. {\tt http://wwwinfo.cern.ch/asd/geant4/geant4.html}

\bibitem{Ahrens:2002hh}
Ahrens J. et al.  [The AMANDA Collaboration] 2002,
%``Search for point sources of high energy neutrinos with AMANDA,''
arXiv:astro-ph/0208006.

\bibitem{Ansorge:1989ba}
Ansorge R.E. et al.  [UA5 Collaboration],
%``Hyperon Production At 200-Gev And 900-Gev Center-Of-Mass Energy,''
Nucl.\ Phys.\ B328, 36.
%%CITATION = NUPHA,B328,36;%%


\bibitem{Battiston:1982su} Battiston R. et al. [UA4 Collaboration] 1982,
%``Measurement Of The Proton - Anti-Proton Elastic And Total Cross-Section At
%A Center-Of-Mass Energy Of 540-Gev,''
Phys.\ Lett.\ B 117, 126.
%%CITATION = PHLTA,B117,126;%%

\bibitem{Barwick} Barwick S. W. et al. [AMANDA Collaboration] 2002, arXiv:astro-ph/0211269.

\bibitem{Bednarek:1997cn}
Bednarek W. and Protheroe R.J. 1997,
%``Gamma rays and neutrinos from the Crab Nebula produced by pulsar
%accelerated nuclei,''
Phys.\ Rev.\ Lett.\ 79, 2616.
%[arXiv:astro-ph/9704186].
%%CITATION = ASTRO-PH 9704186;%%

\bibitem{Ginzburg:sk}  Berezinsky V.S., Bulanov S.V., Dogiel V.A., Ginzburg V.L.,
and Ptuskin V.S. 1990, {\it Astrophysics Of Cosmic Rays,}
North-Holland.

\bibitem{Berezinsky:xp}
Berezinsky V.S., Castagnoli C. and Galeotti P. 1985,
%``High-Energy Neutrino Astronomy With 'Small' Underground Detectors,''
Nuovo Cim.\  8C,  185 [Addendum-ibid.\   8C, 602, 1985].
%%CITATION = NUCIA,8C,185;%%

\bibitem{Bethe} Bethe H.A. and Heitler W. 1934,
%``On the stopping of fast particles
%and on the creation of positive electrons,''
Proc. Roy. Soc. London A146, 83.

%\cite{Bilenky:1998dt}
\bibitem{Bilenky:1998dt}
Bilenky S.M., Giunti C. and Grimus W. 1999,
%``Phenomenology of neutrino oscillations,''
Prog.\ Part.\ Nucl.\ Phys.\  43, 1.
%[arXiv:hep-ph/9812360].
%%CITATION = HEP-PH 9812360;%%

\bibitem{Carroll:1978vq}
Carroll A.S. et al. 1979,
%``Total Cross-Sections Of Pi+-, K+-, P And Anti-P On Protons And Deuterons
%Between 200-Gev/C And 370-Gev/C,''
Phys.\ Lett.\ B80, 423.
%%CITATION = PHLTA,B80,423;%%

\bibitem{Cheng89} Cheng K.S. and Ruderman M. 1989,
%``Period differences between X-ray and very high energy $\gamma$-ray
%observations of accreting X-ray pulsars,''
Astrophys.\ J.\ 337, L77.

\bibitem{Cheng91} Cheng K.S. and Ruderman M. 1991,
%``Stationary accelerators around Keplerian discs of aligned magnetized
%collapsed objects - Pair production and $\gamma$-ray emission,''
Astrophys.\ J.\ 373, 187.

\bibitem{Allen} A. N. Cox (Editor) 2001, {\it Astrophysical
Quantities}, Fourth Edition, Springer-Verlag, New York, p.213-214.

\bibitem{Fesefeldt:yw}
H.~Fesefeldt 1985,
%``The Simulation Of Hadronic Showers: Physics And Applications,''
PITHA-85-02, CERN-DD-EE-81-1, CERN-DD-EE-80-2.

\bibitem{finger} Finger M.H., Wilson R.B. and Harmon B.A. 1996, Astrophys. J. {459},
288.

%\cite{Fukuda:1998mi}
\bibitem{Fukuda:1998mi}
Y.~Fukuda et al.  [Super-Kamiokande Collaboration] 1998,
%``Evidence for oscillation of atmospheric neutrinos,''
Phys.\ Rev.\ Lett.\  { 81}, 1562.
%[arXiv:hep-ex/9807003].
%%CITATION = HEP-EX 9807003;%%


\bibitem{Gaisser:cj}
Gaisser T.K. and Stanev T. 1985,
%``Calculation Of Neutrino Flux From Cygnus X-3,''
Phys.\ Rev.\ Lett.\  { 54}, 2265.
%%CITATION = PRLTA,54,2265;%%

\bibitem{Gaisser:1994yf}
Gaisser T.K., Halzen F. and Stanev T. 1995,
%``Particle astrophysics with high-energy neutrinos,''
Phys.\ Rept.\  { 258}, 173 [Erratum-ibid.\  { 271}, 355 (1996)].
%[arXiv:hep-ph/9410384].
%%CITATION = HEP-PH 9410384;%%

\bibitem{Gaisser:1996qe}
Gaisser T.K., Protheroe R.J. and Stanev T. 1998,
%``Gamma-ray production in supernova remnants,''
Astrophys.\ J.\ { 492}, 219.
%[arXiv:astro-ph/9609044].
%%CITATION = ASTRO-PH 9609044;%%

\bibitem{Ghosh79a} Ghosh P. and Lamb F.K. 1979,
%``Accretion by rotating magnetic neutron stars. II - Radial and vertical
%structure of the transition zone in disc accretion,''
Astrophys.\ J.\ { 232}, 259.



\bibitem{Ghosh79b} Ghosh P. and Lamb F.K. 1979b,
%``Accretion by rotating magnetic neutron stars. III - Accretion torques and
%period changes in pulsating X-ray sources,''
Astrophys.\ J.\ { 234}, 296.



\bibitem{gio} Giovannelli F. and Sabau Graziati L. 1992,
Space Sci. Rev. { 59}, 1.


\bibitem{3EG} Hartman R. C. et al. [EGRET Collaboration],
%``The Third EGRET Catalog of High Energy Gamma Ray Sources,''
Astrophys. J. Suppl. { 123}, 79 (1999).


\bibitem{Hill:2001ds}
Hill  G.C. et al. [the AMANDA Collaboration] 2001,
%``Results from AMANDA,''
arXiv:astro-ph/0106064.
%%CITATION = ASTRO-PH 0106064;%%


\bibitem{janot} Janot-Pacheco E., Motch C. and Mouchet M. 1987, Astron. Astophys. { 177},
91.


%\cite{Karle:2002dv}
\bibitem{Karle:2002dv}
Karle  A. [the IceCube Collaboration] 2002,
%``IceCube: The next generation neutrino telescope at the South Pole,''
arXiv:astro-ph/0209556.
%%CITATION = ASTRO-PH 0209556;%%


\bibitem{Kolb:1985bb}
Kolb E.W., Turner M.S. and Walker T.P. 1985,
%``The Production And Detection Of High-Energy Neutrinos From Cygnus X-3,''
Phys.\ Rev.\ D { 32}, 1145 [Erratum-ibid.\ D { 33}, 859 (1986)].
%%CITATION = PHRVA,D32,1145;%%

%\bibitem{Lipari:hd}
\bibitem{} Lipari P. 1993,
%``Lepton Spectra In The Earth's Atmosphere,''
Astropart.\ Phys.\  { 1}, 195.
%%CITATION = APHYE,1,195;%%


\bibitem{Ranft:fd}
Ranft J. 1995,
%``The Dual Parton Model At Cosmic Ray Energies,''
Phys.\ Rev.\ D { 51}, 64.
%%CITATION = PHRVA,D51,64;%%

%%\cite{Riccobene:2002xk}
\bibitem{Riccobene:2002xk} Riccobene  G. [NEMO Collaboration] 2001,
%``Status of the NEMO project,''
Hamburg DESY - DESY-PROC-2002-01. Published in ``Hamburg 2001,
Methodical aspects of underwater/ice neutrino telescopes'', p.61.

\bibitem{Rhode:es}
Rhode W. et al.  [Frejus Collaboration] 1996,
%``Limits On The Flux Of Very High-Energetic Neutrinos With The Frejus Detector,''
Astropart.\ Phys.\  { 4}, 217.
%%CITATION = APHYE,4,217;%%


\bibitem{Romero:2001wp}
Romero G.E., Kaufman~Bernado M.M., Combi J.A. and Torres D.F.
2001,
%``Variable $\gamma$-ray emission from the Be/X-ray transient A0535+26?,''
Astron.\ Astrophys.\ { 376}, 599.
%[arXiv:astro-ph/0107411].
%%CITATION = ASTRO-PH 0107411;%%



\bibitem {Shap83} Shapiro S.L. and Teukolsky S.A. 1983,
{\it Black Holes, White Dwarfs, and Neutron Stars} (New York: John
Wiley and Sons).
%p.456.

\bibitem{Volkova:sw}
Volkova L.V. 1980,
%``Energy Spectra And Angular Distributions Of Atmospheric Neutrinos,''
Sov.\ J.\ Nucl.\ Phys.\  { 31}, 784  [Yad.\ Fiz.\  { 31}, 1510
(1980)].
%%CITATION = SJNCA,31,784;%%

\end{thebibliography}
\end{document}